\documentclass[a4paper,fleqn]{cas-dc}

\usepackage[authoryear]{natbib}
\usepackage{lineno}
\usepackage{ccicons}

\def\tsc#1{\csdef{#1}{\textsc{\lowercase{#1}}\xspace}}
\tsc{WGM}
\tsc{QE}
\tsc{EP}
\tsc{PMS}
\tsc{BEC}
\tsc{DE}

\begin{document}


\let\WriteBookmarks\relax
\def\floatpagepagefraction{1}
\def\textpagefraction{.001}
\shorttitle{RTApipe framework}
\shortauthors{N. Parmiggiani et~al.}

\title [mode = title]{The RTApipe framework for the gamma-ray real-time analysis software development}                      



\author[1]{N. Parmiggiani}[type=editor,orcid=0000-0002-4535-5329]

\address[1]{INAF/OAS Bologna, Via P. Gobetti 93/3, I-40129 Bologna, Italy.}

\author[1]{A. Bulgarelli}[orcid=0000-0001-6347-0649]

\author[2]{D. Beneventano}[orcid=0000-0001-6616-1753]

\address[2]{Universit\`{a} degli Studi di Modena e Reggio Emilia, DIEF - Via Pietro Vivarelli 10, I-41125 Modena, Italy.}

\author[1]{V. Fioretti}[orcid=0000-0002-6082-5384]

\author[1]{A. Di Piano}[orcid=0000-0002-9894-7491]

\author[1]{L. Baroncelli}[orcid=0000-0002-9215-4992]

\author[1]{A. Addis}[orcid=0000-0002-0886-8045]

\author[3]{M. Tavani}[orcid=0000-0003-2893-1459]

\address[3]{INAF/OAR Roma, Via di Frascati 33, I-00078 Monte Porzio Catone, Italy.}

\author[3,4]{C. Pittori}[orcid=0000-0001-6661-9779]

\address[4]{ASI Space Science Data Center (SSDC), Via del Politecnico snc, I-00133 Roma, Italy.}

\author[5]{I. Oya}[orcid=0000-0002-3881-9324]

\address[5]{CTAO gGmbH, Heidelberg, Germany.}

\begin{abstract}
In the multi-messenger era, coordinating observations between astronomical facilities is mandatory to study transient phenomena (e.g. Gamma-ray bursts) and is achieved by sharing information with the scientific community through networks such as the Gamma-ray Coordinates Network. The facilities usually develop real-time scientific analysis pipelines to detect transient events, alert the astrophysical community, and speed up the reaction time of science alerts received from other observatories. We present in this work the RTApipe framework, designed to facilitate the development of real-time scientific analysis pipelines for present and future gamma-ray observatories. This framework provides pipeline architecture and automatisms, allowing the researchers to focus on the scientific aspects and integrate existing science tools developed with different technologies. The pipelines automatically execute all the configured analyses during the data acquisition. This framework can be interfaced with science alerts networks to perform follow-up analysis of transient events shared by other facilities. The analyses are performed in parallel and can be prioritised. The workload is highly scalable on a cluster of machines. The framework provides the required services using containerisation technology for easy deployment. We present the RTA pipelines developed for the AGILE space mission and the prototype of the SAG system for the ground-based future Cherenkov Telescope Array observatory confirming that the RTApipe framework can be used to successfully develop pipelines for the gamma-ray observatories, both space and ground-based.
\end{abstract}

\begin{keywords}
real-time analysis pipeline \sep 
gamma-ray astronomy \sep 
multi-messenger astronomy \sep 
gamma-ray burst \sep 
gravitational wave
\end{keywords}

\maketitle

\section{Introduction \label{sec:intro}}

We describe in this work the architecture and the key features of the RTApipe, a framework designed to help ground and space-based gamma-ray facilities for the development of real-time analysis (RTA) pipelines in the context of multi-wavelength (MW) and multi-messenger (MM) astronomy.  In recent years, the astrophysical landscape changed after the detection and the follow-up of two crucial transients events. 

The Gamma-Ray Burst (GRB) detected on 17-Aug-2017 by the Fermi Gamma-ray Space Telescope \citep{2018ApJ...861...85A} is the first transient event that confirmed the possibility to perform MM observations of the same physical phenomena. This GRB was detected after the LIGO-Virgo detection of the gravitational wave GW170817, generated during a binary neutron star coalescence \citep{2017PhRvL.119p1101A}. Several MM instruments observed this physical phenomenon during the following days \citep{2017ApJ...848L..12A}. 

The neutrino detected by IceCube (IceCube-170922A) is the second transient event that reinforced the idea of MM observations. Several facilities started an extensive campaign of follow-up with different instruments \citep{2018Sci...361.1378I}. During the campaign, the Fermi-LAT instrument detected an enhanced gamma-ray emission from the blazar TXS 0506+056 in time coincidence with the neutrino. 

These events started the so-called "Multi-Messenger era" \citep{2019NatRP...1..585M}. Now MM observations are key projects for many observatories to study transient phenomena through different "messenger" signals (electromagnetic radiation, gravitational waves - GW, and neutrinos). At the same time, MW astronomy is still important to share information about different wavelengths observations (e.g. infrared, optical, ultraviolet, X-ray, gamma-ray and more).

A science alert is a communication to the astrophysical community that a transient phenomenon occurs in the sky. In today's MW and MM era, the RTA pipelines are mandatory for the observatories that aim to send and receive science alerts about transient events (e.g. GRBs) through networks such as the Gamma-Ray Coordinates Network\footnote{https://gcn.gsfc.nasa.gov} (GCN) and the Astrophysical Multimessenger Observatory Network\footnote{https://www.amon.psu.edu} (AMON).

The GCN \citep{doi:10.1063/1.1361631} is an automated service (without human intervention) that distributes the sky position coordinates of transient phenomena (e.g GRBs, GW and neutrinos) to the community in real-time while the transient event is still ongoing. The GCN network can manage two types of communications: (i) circulars in a human-readable format and (ii) notices to be interpreted by automated software. The observatories can subscribe to the GCN notification channel, selecting the list of facilities from which they want to receive science alerts.

The AMON \citep{2013APh....45...56S,2017ICRC...35..629K} is a consortium of many observatories that aims to search correlations between MM detections (e.g. photons and gravitational waves). In addition, AMON tries to associate sub-threshold signals detected with insufficient statistical significance to become a reliable science alert. The AMON real-time analysis pipeline can correlate these events and find reliable above-threshold detection if more observatories detect a sub-threshold signal simultaneously. When the system finds a reliable detection, it is rapidly distributed to all other observatories and the community to facilitate rapid follow-up. AMON is connected with the GCN network to receive and send science alerts.

The Astronomer's Telegram\footnote{http://www.astronomerstelegram.org} (ATel) is an internet-based service where researchers can publish their short communication about transients phenomena or follow up the results of other observatories. This service is not developed as the GCN network for automated software. Telegrams are available instantly on the website and distributed to subscribers via email. Researchers can periodically check for new telegrams and use the RTA systems' results to search for the same transient event.

We present in this work a framework, called RTApipe, that we designed to facilitate the development of real-time scientific analysis pipelines in the context of gamma-ray observatories. This framework offers several builtin features and allows the researchers to focus on the scientific analyses using existing and new science tools.

We describe in Section \ref{sec:rta_related_works} the solutions adopted by numerous observatories to implement RTA pipelines. We present in Section \ref{sec:rta_usecases} the list of the use cases and the requirements of RTA software that must be satisfied by the RTApipe framework of gamma-ray observatories in the MM and MW context. The RTApipe framework software architecture is described in Section \ref{sec:architecture}. In Section \ref{sec:pipelines} we describe RTA pipelines developed using the RTApipe framework for the AGILE space mission and the Cherenkov Telescope Array (CTA) Observatory. The development of these pipelines shows that the framework can satisfy all the requirements of the gamma-ray observatories.

\section{Real-time analysis pipelines in the multi-messenger era \label{sec:rta_related_works}}

This section describes the RTA software developed by several ground- and space-based observatories in the MM and MW context and shows that they have several common scientific requirements with the gamma-ray observatories. The RTApipe framework shall satisfy these requirements and use cases.

Information on gamma-ray space observatories that implement RTA pipelines are given in the following.

AGILE (Astrorivelatore Gamma ad Immagini LEggero - Light Imager for Gamma-Ray Astrophysics) is a scientific mission of the Italian Space Agency (ASI) that performs X-ray and gamma-ray observations and was launched on 23rd Apr 2007 \citep{2008NIMPA.588...52T, 2009A&A...502..995T}. The AGILE payload consists of the Silicon Tracker (ST), the SuperAGILE X-ray detector, the CsI(Tl) Mini-Calorimeter (MCAL), and an AntiCoincidence System (ACS). The combination of ST, MCAL, and ACS forms the Gamma-Ray Imaging Detector (GRID). We developed several RTA pipelines, described in Section \ref{sec:agile_rta}, to detect transient events (e.g. GRBs) and react to external science alerts received by the GCN network.

The Swift space mission \citep{2004ApJ...611.1005G} is an MW observatory for GRB astronomy, renamed in 2019 to Neil Gehrels Swift Observatory. It is an autonomous rapid-slewing satellite to study transient astronomy and observes more than 100 GRBs/year performing a detailed X-ray and UV/optical afterglow follow-up in timescales ranging from 1 minute to several days after the GRB. The results generated in real-time onboard by the Burst Alert Telescope (BAT \cite{2005SSRv..120..143B}) instrument are transmitted to the ground processing software that sent them to the GCN network. 
In addition to the analyses performed on-board to detect GRBs in the Swift/BAT data, the Swift team developed a new pipeline to save the event-mode data, usually stored in a ring buffer overwritten after it reaches a maximum size. These data can be saved on-demand to follow up triggers from external instruments \citep{Tohuvavohu_2020} and are rapidly sent to the ground to be analyzed with powerful analyses methods that improve the detection capability over the on-board detection system. This new feature can be used mainly to analyze subthreshold GRBs and search coincidences between GWs and GRBs

The Fermi Gamma-Ray Space Telescope \citep{2009ApJ...697.1071A} was launched on 2008 June 11, in a 565 km orbit with an inclination of 25.6 deg. The payload comprises two science instruments, the Large Area Telescope (LAT, \cite{Ackermann_2012}) and the Gamma-Ray Burst Monitor (GBM, \cite{2009ApJ...702..791M}). The GBM on-board flight software detects and localizes GRBs.  This is followed by improved ground-based localizations, and prompt but not real-time detection of fainter GRBs. When a GRB is detected, the pipelines send a circular and a machine-readable notice to the GCN network. The Fermi team developed an Automated Science Processing (ASP) system that detects short and long transient phenomena in the Fermi/LAT processed data (Chapter 4 of the book \cite{2012amld.book.....W}) and sends notices to the GCN network. The ASP software can also react to notices received by other facilities.

Several ground-based facilities for studying gamma-rays (e.g. MAGIC, H.E.S.S, HAWC, and more) implemented RTA pipelines and CTA will have an RTA system.

The Major Atmospheric Gamma-ray Imaging Cherenkov Telescopes (MAGIC, \cite{TRIDON2010437}) experiment is composed of two 17 m diameter mirror imaging atmospheric Cherenkov telescope (IACT, \cite{Krennrich_2009}), called MAGIC-I and MAGIC-II. The two telescopes can observe the high energy gamma-ray sky independently or in stereoscopic mode. The MAGIC team developed an automatic alert system \citep{2019ICRC...36..633B} that receives MM science alerts from the GCN network and follow up the transient events. 

The High Energy Stereoscopic System (H.E.S.S.\footnote{http://www.mpi- hd.mpg.de/hfm/HESS/}) is an array of five IACT telescopes dedicated to the observation of very-high-energy (VHE) gamma rays and located in the Khomas Highland of Namibia. The H.E.S.S observatory has a central data acquisition system (DAQ, \cite{2014APh....54...67B}) that executes real-time analyses to produce preliminary results (e.g. calibrated camera images, significance maps, and $\theta^2$ plots) that can be visualised through a graphical user interface (GUI). In addition, the DAQ system manages the external science alerts about GRBs received by the GCN network.

CTA \citep{2011ExA....32..193A} is an initiative to build the next generation of ground-based gamma-ray astronomy made by dozens of IACT telescopes that provide an unprecedented sensitivity to detect transient events \citep{2019ICRC...36..673F}. CTA will be provided with an RTA software, called Science Alert Generation (SAG, \cite{Bulgarelli:20210h}), to perform analysis as soon as the telescopes acquire data. This software is one of the sub-systems of the Array Control and Data Acquisition System (ACADA, \cite{oya:hal-02421340}) of the CTA observatory. We developed a prototype for the SAG-sci RTA pipeline that is one of the SAG sub-systems. The SAG-sci aims to perform scientific analyses on the acquired data and generate science alerts (Section \ref{sec:cta_rta_pipe}).

In the MM context, ground-based observatories built to detect neutrinos and GWs implemented RTA software systems to send and receive science alerts.

The IceCube Neutrino Observatory \citep{2017JInst..12P3012A} is a cubic-kilometre detector built into the ice at the South Pole to detect neutrinos and operative since 2011. The IceCube automated system \citep{2017APh....92...30A} performs analyses on data acquired by the instrument to detect neutrinos events and determine if science alerts should be generated and shared with other observatories through the AMON and GCN networks. IceCube collaborates with MM observatories because the detection of neutrinos has a crucial role in notifying other observatories about the localisation of the transient events.

The Laser Interferometer Gravitational-Wave Observatory (LIGO \cite{Abbott_2009}) is constituted of three specialised Michelson interferometers. The three detectors are built-in separated sites to reject instrumental and environmental perturbations into the data and perform a triangulation between the different locations to localise the source of the GWs. The low-latency detection of GWs is critical to trigger the follow-up of GW events by other observatories that can detect prompt GRBs and neutrinos within seconds \citep{PhysRevD.95.042001}. Therefore, the LIGO team developed several automated pipelines to process the data acquired by the detectors, obtain low-latency results ($\lesssim$ 1 minute), and share the GWs with the scientific community through the GCN network.

\section{Use Cases and Requirements\label{sec:rta_usecases}} 

This section describes the use cases and requirements in the context of gamma-ray astronomy that the RTApipe framework shall satisfy. The starting points are the requirements of the CTA SAG and the know-how from the AGILE mission acquired by the AGILE team in the last 15 years of scientific operations in the context of MM and MW astrophysics. The generic RTA pipeline developed with the framework shall be able to achieve two main use cases:

\begin{enumerate}
  \item execute periodical scientific analysis, searching for transient events and promptly generate science alerts to share the information with the community;
  \item react to external science alerts and start scientific analysis on available data or wait that the requested data become available.
\end{enumerate}

The RTApipe framework aims to facilitate researchers to quickly develop RTA pipelines for gamma-ray observatories that satisfy all the requirements.

The main requirements of the framework derived from the CTA SAG requirements and from the know-how acquired with the AGILE RTA system are:

\begin{enumerate}
    \item\label{req:diff_observatories} it shall be usable in different contexts (e.g. space or ground-based, single telescope or array of telescopes);
    \item\label{req:diff_array} it shall analyse data acquired by several arrays of telescopes that are observing different sky regions;
    \item\label{req:parallel_analysis} it shall execute a great number of analyses in parallel;
    \item\label{req:task_priority} it shall manage queues and priorities between processes to optimise the available resources;
    \item\label{req:diff_power_need} it shall be easily scalable on more computing resources without change the software implementation but only the configurations;
    \item\label{req:diff_workflow} it shall allow the configuration manager to configure different analysis workflows;
    \item\label{req:diff_alert_type} it shall perform different type of analyses for each type of science alert or instrument that triggers them (e.g. GW, neutrino, Fermi/GBM etc.);
    \item\label{req:diff_st} it shall be configurable with many different science tools and allow the configuration manager to update their configurations;
    \item\label{req:diff_time_scales} it shall perform analyses at different time scales (from seconds to hours) using a given science tools;
    \item\label{req:deployability} it shall be easily deployable with all its dependencies.
    \item\label{req:streaming} it shall analyse data arriving in streaming (event by event) or in batch mode (chunks of events).
    \item\label{req:replay} it shall analyse new data as they arrives or perform analyses on data already present in the archive.
\end{enumerate}

The following section describes the framework architecture and the solutions adopted to fulfil the requirements. 

\section{Framework architecture \label{sec:architecture}}

This section describes the architecture of the RTApipe framework and all its components: Data Model (DM), Science Logic, Pipeline Manager, and Task Manager. The framework must be integrated using external interfaces with additional components: data sources, science tools, analysis results. The DM defines the framework's data structure, the entities and the relationships between them. The DM is very flexible and allows the framework to implement pipelines for different contexts (e.g space missions or arrays of telescopes). The Science Logic implements the rules that execute the analyses and satisfy the use cases and the requirements of a gamma-ray observatory. Finally, the Pipeline Manager and the Task Manager execute the analyses and manage priority and queues between processes.

\begin{figure*}[!htb]
	\centering
	  \includegraphics[width=\textwidth]{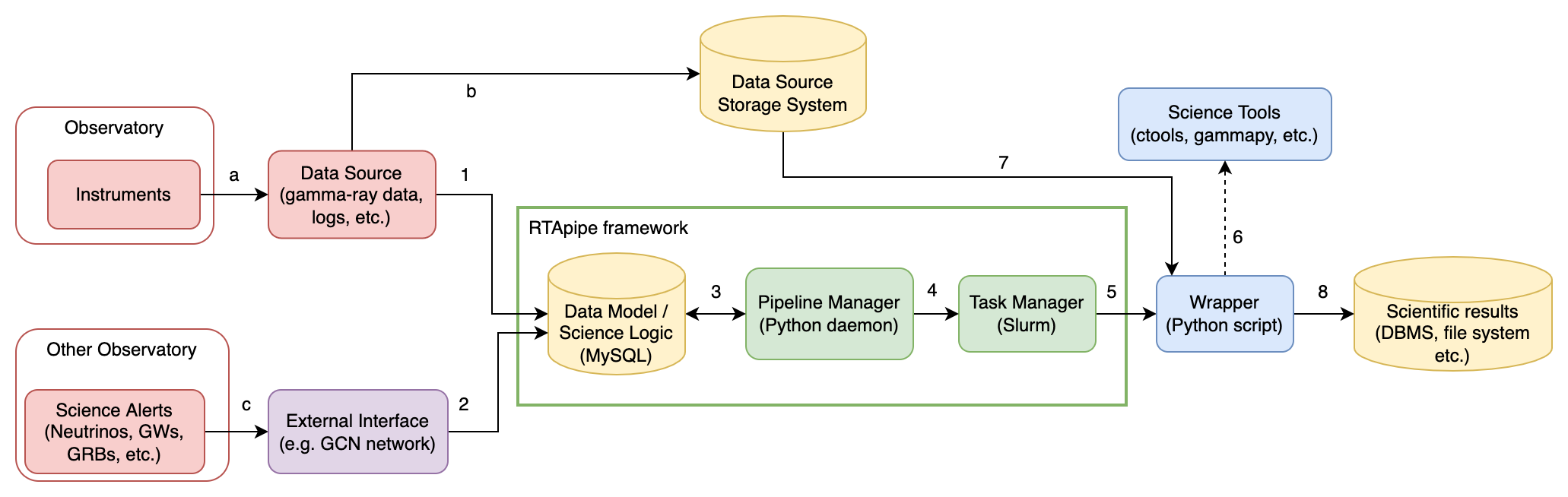}
	\caption{High-level schema of the RTApipe framework software architecture. The schema shows all the components and external entities that are interfaced with the framework. The numbers from 1 to 8 indicate the workflow starting from the input data to the scientific results. The dashed line represents the software dependency between the wrappers and the science tools. The letters indicate actions that are not managed by the RTApipe framework.}
	\label{fig:architecture}
\end{figure*}

Figure \ref{fig:architecture} shows a high-level overview of the framework's architecture and its components. The RTA pipeline has two main input types. The firsts are the data types generated by the instruments (e.g. gamma-ray data, logs, auxiliaries, etc.). These data are managed by a Data Source software component (step a) that stores them in a generic storage system (step b) and notifies the RTApipe framework of the arrival of new data (step 1). The seconds are the science alerts generated by other observatories (step c) and received through external interfaces such as the GCN network (step 2). These two inputs start the execution of the Science Logic rules, which generate new analyses described by the DM. The Pipeline Manager periodically queries the DM to check if new analyses are ready to be processed (step 3), then submits them to the Task Manager (step 4). The Task Manager executes the Python wrapper that interfaces the required science tools (step 5) on the available computing resources. The wrapper uses the science tools to execute the analysis (step 6) and reads the needed data from the Data Source storage system (step 7). Finally, the wrapper saves the results obtained by the analyses in the chosen storage system (DMBS, file system, etc.) outside the RTApipe framework (step 8).

\subsection{Data Sources}\label{sec:datasources}

The RTA framework does not manage the data directly but has an interface with them. The data can be stored in a dedicated storage system by the Data Source. The RTA framework is interfaced with the Data Source to receive a notification when new data are available for the analyses. When the pipeline receives a notification from the Data Source, it automatically starts the new analyses that query the storage system to retrieve the required data.

\subsection{Data Model}\label{subsec:data_model}

\begin{figure*}[!htb]
	\centering
	  \includegraphics[width=0.8\linewidth]{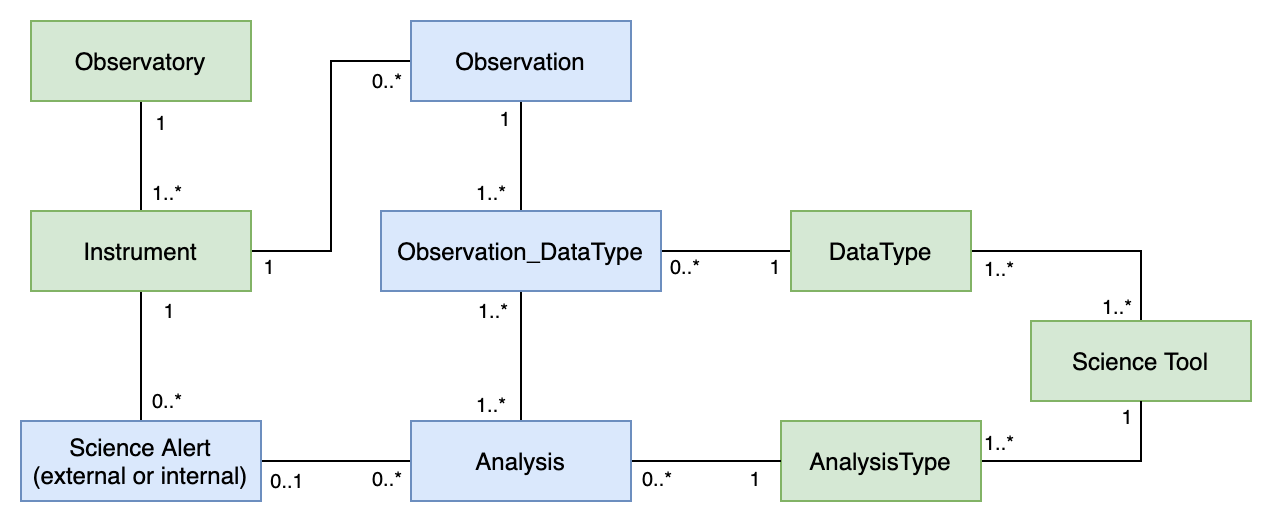}
	\caption{High-level schema of the data model displayer with an UML class diagram. Static entities and relationships are in green, while dynamic entities and relationships are in blue. The schema also shows the multiplicity between entities to show the lower and upper bound on the number of associated entities. The "*" symbols stand for no upper limit to this number. }
	\label{fig:datamodel}
\end{figure*}

The DM represents all the entities involved in an RTA pipeline for different types of gamma-ray observatories (e.g. satellites or ground telescopes): Data Sources, Instruments, Observations, Analysis, Science Tools, and more. Figure \ref{fig:datamodel} shows a high-level schema of the DM with the main entities included in the RTApipe framework. The configuration managers of the pipeline can configure the DM to implement the desired workflow. The DM comprises a static part (showed in green in Figure \ref{fig:datamodel}) which is configured before the start of the analyses, and a dynamic part (showed with blue in Figure \ref{fig:datamodel}) that the RTA pipeline updates during operations.

We implemented the DM using the MySQL\footnote{https://www.mysql.com} relational database management system. The complete DM consists of 43 entities and 57 relationships between them.

The main DM static entities are:
\begin{itemize}
    \item Observatory: contains the list of observatories managed by the system that can send science alerts or produce data to be analysed. This table includes the observatory for which the pipeline is being developed.
    \item Instrument: an observatory can have more than one instrument. These instruments can generate data if they are part of the observatory that is developing the pipeline or send science alerts if they are part of external observatories.  
    \item Data Type: the data type describes the data (i.e. model and format) generated by the observatory's instruments and used for the scientific analyses;
    \item Data Source: the system that manages and stores data types;
    \item Science Tools: the scientific software packages used for the analyses of scientific data;
    \item Analysis Type: the configuration of the analyses that define the usage of the science tools (e.g. creating sky maps with a time window of 100 sec).
\end{itemize}

The main DM dynamic entities are:
\begin{itemize}
    \item Observation: contains the main parameters that describe the current observation that the observatory is performing, between them there are the pointing coordinates, the target source, the observing mode and others;
    \item Observation\_DataType: implements the N, N relationship between the observations and the data types produced during the observations;
    \item Analysis: this entity is updated during the operations. It contains technical parameters (e.g. processing time, status, etc.) and references to several statics entities (e.g. the analysis type and the instruments that generated the data). The information contained in this entity can be used to reconstruct the analysis details;
    \item Science Alert: contains the information about science alerts received through the GCN network (external) or generated by the observatory (internal).
\end{itemize}

\subsection{Science Logic}\label{subsec:bus_logic}

The Science Logic defines the set of rules used by the pipeline to know \textit{When} and \textit{How} execute the configured analyses during operations and that are described by the Analysis Type entity. The RTApipe implements all the rules to satisfy the use cases and requirements described in Section \ref{sec:rta_usecases}. We implemented the Science Logic rules with 12 MySQL triggers running in the same database as the DM. The triggers are stored procedures invoked automatically by the database before or after an event (insert, update, and delete) that occurs in the associated table. For example, a trigger can be invoked automatically before a new row is inserted into a table. MySQL triggers can execute Science Logic algorithms with low latency and consistency. The triggers are used to implement the desired analysis workflows. 

This component retrieves the analysis configuration from the static part of the DM and then updates the dynamic part. The set of rules, defined in advance, allows the pipeline to operate in real-time without human intervention. We developed these rules to perform new analyses when one of the following two scenarios occurs:

\begin{itemize}
    \item new data acquired by the instruments are available for configured analyses;
    \item a new science alert (internal or external)  is received and inserted in the DM.
\end{itemize}

The first scenario occurs when a Data Source notifies the pipeline that new data of one or more data types arrived. The configuration manager can configure a minimum time window of data required to start the analyses (e.g. 100 sec, 1000 sec etc.). If no time window is configured, the pipeline performs the analyses at each update. The pipeline trigger checks if inside the Analysis Type table there are analyses configured for the data type arrived. If yes, the trigger creates new analyses that can be in two states: i) 'pending for data' when not all required data are available and ii) 'runnable' when all analyses requirements are satisfied, and the system can execute them. When new data are available, the trigger checks if analyses in the 'pending for data' state can be updated in the 'runnable' state. Once in the 'runnable' state, the analyses are ready to be submitted to the Task Manager by the Pipeline Manager component.

The second scenario occurs when the pipeline receives an internal or external science alert and starts the follow-up of the transient event. The trigger involved in this scenario checks if the Analysis Type entity contains analyses configured for the science alert type received or the instrument that sent it. If yes, it creates the analyses with a 'pending for data' state. The trigger checks if the data archive contains the data in the required time window for the analyses and sets the analyses status to 'runnable'; otherwise, the pipeline waits until the required data arrives. The RTA pipeline creates the analysis in both scenarios starting from the configuration defined in the Analysis Type entity. 

The configuration manager can exploit the flexibility of the RTApipe Science Logic to:

\begin{enumerate}
    \item configure a group of data types required to start a specific analysis. The trigger checks for the data availability or wait until the needed data become available;
    \item perform the analyses integrating in time for the entire duration of an observation;
    \item configure a flexible analysis time window without a fixed size that starts at the end to the last analysis time window;
    \item set the time window for the the analyses that can be a single fixed time window or a large time window divided into various bins (e.g. perform analysis inside a 2000 s time window with bins of 100 s);
    \item configure a list of analyses with predefined time windows centred on the event time given by the science alert (e.g. even time $\pm 100 s$); 
    \item  configure a list of analyses that start only when a science alert is received from a particular instrument (e.g. from IceCube);
    \item configure a partial analysis that starts if data coverage is not complete. When the entire data arrives, another analysis starts with full data coverage; if the partial analysis is still running, it is stopped to release resources for the full coverage analysis;
    \item configure the sky position coordinates used for the analyses as the position of a known gamma-ray source, the centre of the observation field of view, or a list of coordinates defined with a grid of points (HEALPix, \cite{Gorski_2005}) useful to analyse sources with large error regions (e.g. GWs);
    \item define processes, called post-analysis, that is executed when a list of associated analyses ends without error (e.g. import results on a database or merge results from several analyses); 
    \item define the priority between different analysis types and configure the scheduler with queues. High-priority analysis can suspend low-priority analysis;
    \item disable or enable the analysis types by changing a flag into the DM.
\end{enumerate}

The pipeline configuration manger configures these parameters and analysis types. Then, the triggers manage the analysis automatically during the operations.

\subsection{Pipeline Manager \label{subsec:pipeline_manager}}

The main task of the Pipeline Manager component is to execute the analyses generated by the Science Logic rules on the available computational resources. The Pipeline Manager can manage several pipelines (e.g. one for each detector onboard a satellite). Each pipeline has its own DM, and the Pipeline Manager reads the connection information to these DMs from a configuration file. 

\begin{figure*}[!htb]
	\centering
	  \includegraphics[width=0.9\linewidth]{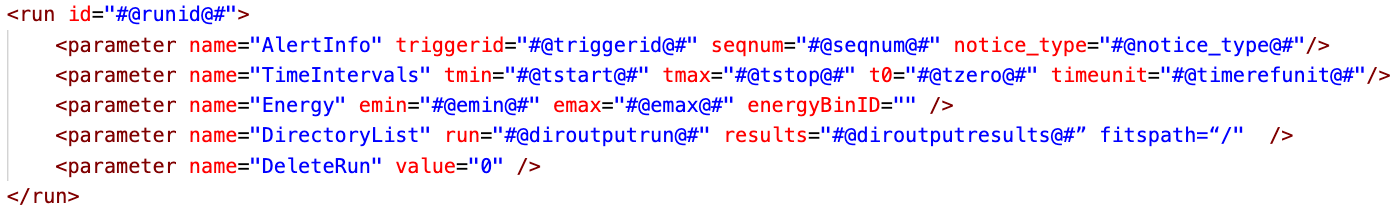}
	\caption{Example of an XML template for the configuration of the analysis executed after the arrival of an external science alert. The XML contains special tags identified with "$\#@@\#$" updated at runtime by the Pipeline Manager.}
	\label{fig:xml_example}
\end{figure*}

The Pipeline Manager is written in Python and periodically queries the DMs of all configured pipelines to check if the triggers have created new analyses in the 'running' state ready to be executed. When the Pipeline Manager finds new analyses, it creates a new directory in the file system, queries the DM to obtain all the information necessary to execute these analyses (time window, science tool configurations, and more) and saves this information inside a file written using the Extensible Markup Language (XML). Then, the pipeline executes the analyses with the configured science tools and required parameters using the information contained in the XML file. The DM contains a template of XML configuration for each science tool (Figure \ref{fig:xml_example}). The template has a dynamic part that contains parameters with special tags that are replaced at run time by the Pipeline Manager with real information and a static part customised to include fixed parameters for a specific science tool. Finally, the Pipeline Manager submits the configured analysis to the Task Manager.

\subsection{Task Manager}\label{sec:task_manager}

Since the number of analyses executed simultaneously can be very high, the Pipeline Manager does not process them directly but submits the analyses to a Task Manager component. The Task Manager is a job scheduler that performs the analyses following the priorities defined in the DM. This component executes the analyses in parallel, manages different jobs queues and optimises the computational resources. The Task Manager can temporarily suspend low-priority running analyses to execute higher priority analyses (e.g. when an external alert is received).

We implemented the Task Manager using Slurm\footnote{https://slurm.schedmd.com}, an open-source, fault-tolerant, and highly scalable job scheduling system for large and small Linux clusters. Slurm can manage thousands of jobs and can easily scale from one to hundreds of servers by modifying a few lines in the Slurm configuration file.  Increasing computing power does not need to change the DM or the Science Logic. If Slurm is configured in a cluster of multiple nodes, it can also manage the nodes' failure, keeping the pipeline operational. Another important feature of Slurm is the capability to log inside a MySQL database all the information about every process. The logging service is helpful to monitor the pipeline in real-time and visualise statistics about the execution of tasks. 

\subsection{Science Tools}\label{sec:science_tools}

Researchers develop science tools to generate scientific results (e.g. light curves, sky maps, detections, etc.) from input data. Usually, the science tools are updated during the project lifetime by different development teams. Hence it is crucial to have the flexibility to change the science tools inside the pipeline, maintaining the same interfaces. The configuration manager configures the science tools inside the DM before the start of the observation. The RTApipe framework can manage different science tools and perform different analyses using the same science tool. In addition, the pipeline can execute analyses using science tools developed with several programming languages or software libraries using a Python wrapper as interface.

\subsection{Analysis Results}\label{sec:ana_results}

The RTApipe framework does not manage the analysis results directly. The science tool wrappers can save their results in different storage systems (e.g. database, file system, and cloud services), and this behaviour has no impact on the pipeline architecture. The pipeline offers a feature to execute processes that manage the results after the analysis completion (e.g. import the results into a database).

\subsection{External Interfaces}\label{subsec:external_interfaces}

The RTApipe framework requires the integration with components not provided within the framework through external interfaces:

\begin{enumerate}
    \item the interface between the science tools and the RTApipe framework implemented with science tools wrappers;
    \item the interface between the observatory Supervisory Control System of the observatory and the RTApipe framework to inform the pipeline when an observation begins or ends;
    \item the interface with data sources to receive notifications when new data are available;
    \item the interface with other observatories to receive external science alerts through different networks such as the GCN or AMON. Usually, a Transients Handler systems manages these interfaces to send and receive science alerts.

\end{enumerate}

\subsection{Science Tool Wrapper}\label{subsec:wrapper}

The wrapper is a Python script that interfaces the science tools with the RTA pipeline reading the configurations required to execute the analysis from the XML file written by the Pipeline Manager. For this reason, the wrapper is an essential component of the RTApipe framework to interface several science tools with the pipeline without changing the system architecture. Each science tool requires the development of a specific wrapper.

\subsection{Framework key features} \label{sec:rta_key_features}

The solutions adopted for the architecture of the RTApipe framework satisfy the requirements described in Section \ref{sec:rta_usecases} providing a list of valuable features:

\begin{enumerate}
    \item \textit{Parallel analysis and task priority.} The pipelines developed with the RTApipe framework can perform analyses in parallel, exploiting the Slurm capabilities. If the computing resources are not enough to run all analyses, it is possible to prioritise them. It is also possible to configure job queues and manage the priorities between different queues (e.g. a queue for each data source).
    \item \textit{Flexibility and scalability} The framework does not require the usage of a particular input data format or output storage system because they are managed by the wrapper developed for each science tool. The pipeline can work with any science tool developed with different programming languages and allows the configuration manager to configure new ones when required.
    \item \textit{Pipeline Monitoring} The Task Manager logs in a MySQL database all the information related to the execution of the analyses. It is possible to use these monitoring data to calculate several metrics, such as the average time needed to perform a specific analysis or get the list of analyses ended with errors to search for bugs. Since this information is updated in real-time, a GUI can periodically query the database and shows an overview of the system processes.
    \item \textit{Containerisation and easy setup} All the software, the services and dependencies of the RTApipe framework (MySQL, Slurm, Python, and more) can be installed and configured inside a Singularity\footnote{https://sylabs.io/docs/} container if necessary. If the services are already present inside the IT infrastructure, the RTApipe can be installed as self-contained software. This system allows a quick and straightforward setup on different types of hardware and Operating Systems. The use of containers is also helpful to integrate the pipeline into a continuous integration workflow.
\end{enumerate}

\section{Pipelines developed with this framework} \label{sec:pipelines}

This section describes the pipelines developed using the RTApipe framework. These pipelines are developed in different contexts and shows how the framework satisfies the requirements.

\subsection{AGILE real-time analysis software} \label{sec:agile_rta}

\begin{figure*}[!htb]
	\centering
	  \includegraphics[width=\linewidth]{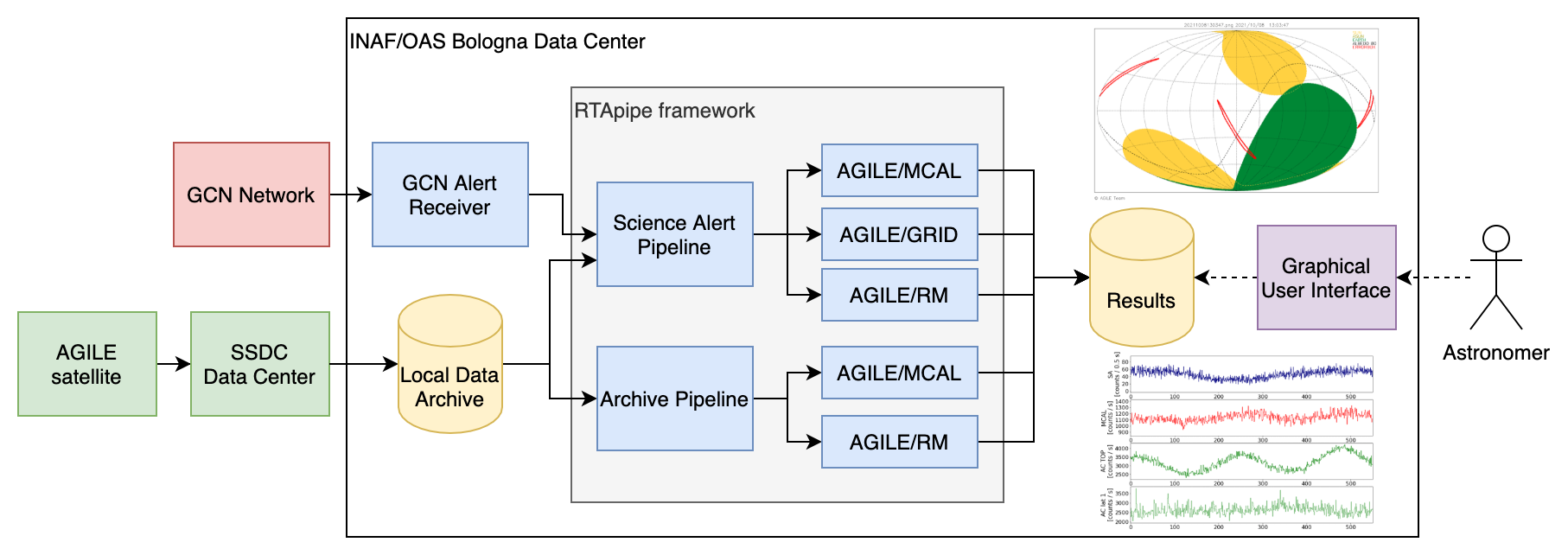}
	\caption{Architecture of the RTA system of the AGILE space mission developed with the RTApipe framework. It consists of three main pipelines that react to science alerts or analyse the instruments' data as soon as they are received.}
	\label{fig:agile_rta_architecture}
\end{figure*}

Figure \ref{fig:agile_rta_architecture} shows the architecture of three RTA pipelines, described in detail in \cite{Parmiggiani:20214o}, that are part of the AGILE RTA system \citep{2019ExA....48..199B}. We developed these pipelines using the RTApipe framework. The inputs for the RTA pipelines are: (i) the science alerts received by the GCN network, managed by the GCN Alert Receiver component and (ii) the data acquired by the detectors onboard the AGILE satellite, downlinked to the ground and sent to the Space Science Data Center\footnote{https://www.ssdc.asi.it} (SSDC, Rome, Italy). The SSDC performs the reduction, archiving, and distribution of data to the INAF/OAS Bologna (Italy) data center local archive through the Telemetry Preprocessing System (TMPPS, \cite{10.1117/12.789338}). We developed two types of pipelines: the Science Alert Pipelines and the Archive Pipelines. 

The Archive Pipelines analyse data acquired by the satellite detectors as soon as they are available in the local archive to detect transient events and share them with the community of observatories. The first archive pipeline is the AGILE/MCAL pipeline that analyses data acquired by this detector. In this case, the pipeline's triggers (Section \ref{subsec:bus_logic}) are executed automatically and can create new analyses based on the configuration. The Pipeline Manager submits these analyses to the Task Manager that executes them in parallel to obtain scientific results. The analyses are focused on the detection of GRBs and Terrestrial Gamma-ray Flashes (TGF, \cite{2014JGRA..119.1337M}) events. When this pipeline detects a GRB, it automatically prepares and sends a notice to the GCN network to inform other facilities in real-time without human intervention. Since we developed this feature inside the AGILE/MCAL pipelines (May 2019), it sent more than 50 notices of GRBs to the GCN network\footnote{https://gcn.gsfc.nasa.gov/agile\_mcal.html}. The second archive pipeline is the AGILE/RM pipeline developed to search for GRBs and solar flares in the detectors' ratemeters (RM) data. The RMs represent the acquisition event rate of a detector over time (e.g. the number of photons or particles detected per second). An automated algorithm detects anomalies in the RM light curve using predefined thresholds when a transient event increases the count rate. The pipeline sends an email to the AGILE team when a transient event is detected. 

We designed Science Alert Pipelines to react to internal or external science alerts. When the Archive Pipelines detect transient phenomena, they generate internal alerts. In this case, the pipeline searches for counterpart detection in other detectors (e.g. a science alert received by AGILE/MCAL triggers analysis on AGILE/GRID data). On the other hand, external science alerts are received from other facilities (e.g. Fermi, LIGO and IceCube) through the GCN network with notice format. When an external science alert is received, the pipeline informs the AGILE team via email and executes more than 100 processes with 20 different science tools to analyse the AGILE/GRID, AGILE/MCAL, and AGILE/RM data. The system is flexible enough to implement Deep Learning techniques for the GRB detection \citep{2021ApJ...914...67P}. Slurm manages the execution of these analyses (Section \ref{sec:task_manager}). The RTApipe framework features are essential to satisfy the AGILE team requirements to follow-up external science alerts and generate quick results in seconds or minutes (depending on the task) since the notice arrival. 

Both Science Alert and Archive Pipelines store the results into a MySQL database and in the file system. The team can visualise them through a password-protected web GUI that queries the database and shows the images present in the file system. The GUI is installed within a LAMP environment (Linux, Apache, MySQL, PHP), and the layout is implemented with the Bootstrap\footnote{https://getbootstrap.com} framework enabling a responsive behaviour compatible with smartphones and tablets. The plots are implemented with the Plotly.js\footnote{https://plotly.com/javascript/} framework. The used programming languages are Javascript and PHP, together with the HTML markup language and the CSS style sheet language.

The astronomer on-duty, a team member responsible for monitoring the results of all AGILE automated pipelines, must perform more detailed manual analyses to investigate the phenomena when the GUI displays a result. After the investigation, the astronomer on-duty can decide to communicate the results to the community to trigger follow-up observations by other facilities. To facilitate the astronomer on-duty in this process, the Science Alert Pipeline prepares two templates of the GCN circulars' texts that can be verified and sent to the GCN network responding to the received science alerts.

In addition to the communication with the GCN network, the AGILE team published several papers using the results obtained with the RTA pipelines as a support system during data analysis. The researchers can manually inject science alerts to execute analyses offline on the archived data.

\subsection{CTA high-level real-time analysis prototype} \label{sec:cta_rta_pipe}

The SAG system of the CTA Observatory, introduced in Section \ref{sec:rta_related_works}, is divided into different components. One of these components is the SAG-sci or high-level analysis pipeline that receives in input the reconstructed event list (EVT3) and produces high-level scientific results (e.g. sky maps, light curves). In addition, the pipeline aims to detect transient events and generate candidate science alerts that will be validated and eventually shared with the community by another ACADA sub-system called Transients Handler.

The list of main SAG requirements and functionalities is ìdescribed in \cite{Bulgarelli:20210h}. These requirements drove the use of the RTApipe framework to develop this prototype: 

\begin{itemize}
    
    \item the SAG-sci is a software system that shall analyse CTA data in real-time during the observation;
    \item the computational resources shall be optimised due to the software deployment in the on-site data center;
    \item the analyses shall analyse data and detect candidate science alerts within 5 seconds since the event list is received;
    \item the software shall be able to analyse multiple observations in parallel due to CTA capability of operating up to 8 sub-arrays simultaneously;
    \item the analysis shall search for transient phenomena on different timescales, from 10 seconds to 30 minutes using different science tools;
    \item the sensitivity of the analysis shall not be worse than the offline analysis by more than a factor of 2.
    
\end{itemize}

We developed a prototype of the SAG-sci pipeline using the RTApipe framework to prove that the framework satisfies the CTA use cases and requirements. In addition, this prototype can be used as a starting point to develop the final SAG-sci software system.

\begin{figure*}[!htb]
	\centering
	  \includegraphics[width=0.9\textwidth]{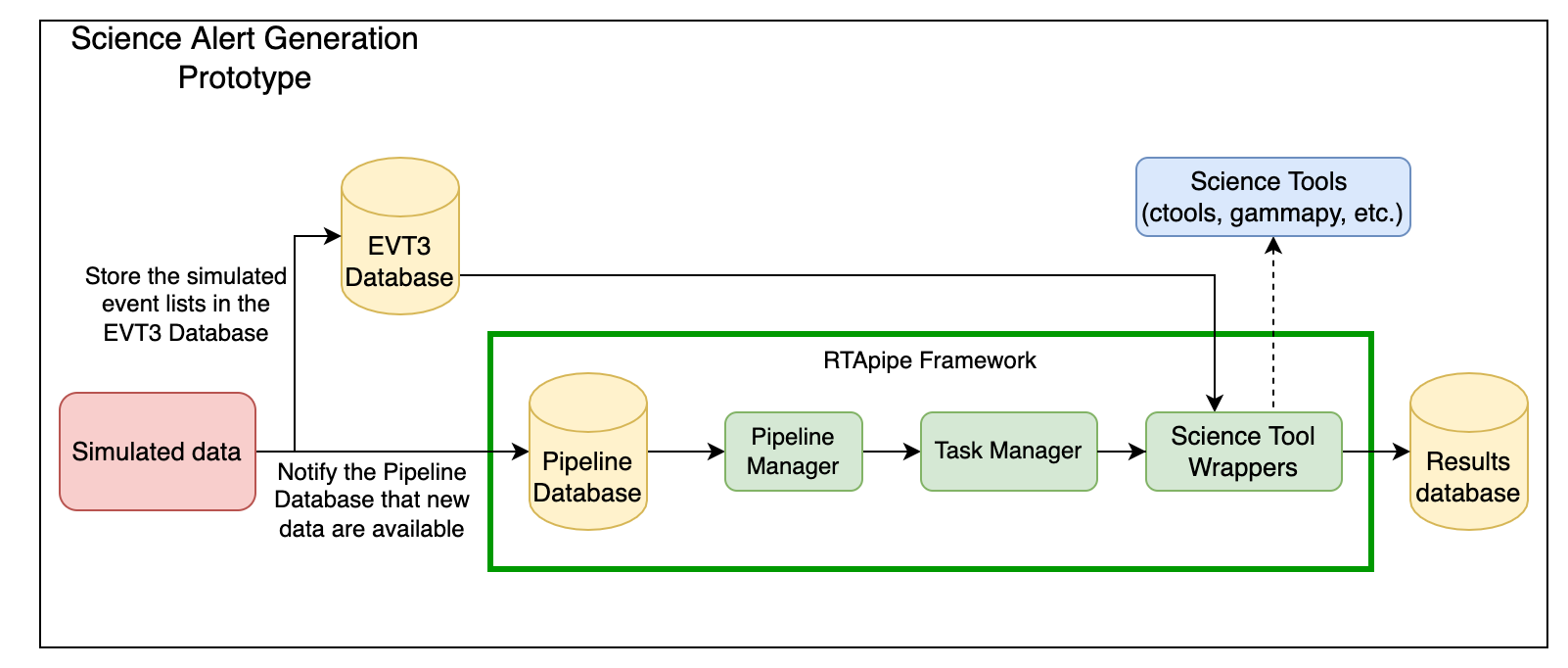}
	\caption{High-level functional schema of the SAG-sci pipeline prototype for the CTA real-time scientific analysis. This pipeline executes the analysis using the configured science tools through the wrappers when simulated data are sent to the pipeline database. The dashed line represents the software dependency between the wrappers and the science tools. The results are stored in a MySQL database.}
	\label{fig:cta_rta_pipe_schema}
\end{figure*}

The diagram in Figure \ref{fig:cta_rta_pipe_schema} shows a high-level functional schema of the SAG-sci prototype. The RTApipe framework is the orchestrator of the pipeline. The prototype receives a simulated event list and executes a list of pre-configured analyses with different science tools. When new simulated events are inserted into the EVT3 database, the EVT3 time index inside the DM is updated, and the Science Logic triggers create the analyses that the Pipeline Manager submits to the Task Manager. We developed several Python wrappers to interface the SAG-sci prototypes with different science tools (ctools\footnote{http://cta.irap.omp.eu/ctools/}, gammapy\footnote{https://gammapy.org}, and ad hoc analysis software) \citep{DiPiano:2021+e}.  

The users can visualise the prototype results, stored in a MySQL database, through a web GUI developed as part of the prototype. We implemented this GUI just for engineering purposes starting from the AGILE RTA GUI and adding some features such as the d3-celestial library\footnote{https://github.com/ofrohn/d3-celestial} to plot the interactive sky map in Galactic coordinates. This GUI is not meant to be proposed as a solution for the CTA project.

\section{Conclusions}

This work describes the RTApipe framework designed to facilitate real-time analysis pipelines development for gamma-ray observatories. This framework has several key features that satisfy the scientific analysis requirements in the multi-wavelength and multi-messenger era. In this context, observatories share their results about transient events detected with different electromagnetic wavelengths or messenger signals (electromagnetic radiations, neutrinos and gravitational waves), sending science alerts through networks (e.g. the GCN). Science alerts are received and managed by automated software systems developed by the observatories to react rapidly to transient events and search for their counterparts. Sharing these science alerts is mandatory to study transient events using the results obtained by different messenger or wavelength detectors.

Two main scenarios can be implemented using this framework: (i) perform prompt scientific analyses on the data acquired by the instruments searching for transient sources, and (ii) react to science alerts received by other facilities through communication networks and start automated analyses to follow up the transient phenomena. The RTApipe framework has several key features: (i) it can manage several analyses in parallel and the priority between them, optimising the available computational resources, (ii) the configuration manager can straightforwardly configure the system to perform analyses with new science tools, providing high flexibility for future software updates and maintenance, and (iii) all the services and the software component of this framework are installed and provided within a Singularity container that is easy to deploy in several environments.  
 
The RTA pipelines developed for the AGILE space mission and the prototype of the SAG system for the CTA observatory confirm the capabilities of this framework to satisfy the requirements. Consequently, we conclude that the RTApipe framework can be used to successfully develop RTA pipelines in MM and MW context for the gamma-ray observatories, both space and ground-based.

\section*{Acknowledgements}

The AGILE Mission is funded by the Italian Space Agency (ASI) with scientific and programmatic participation by the Italian National Institute for Astrophysics (INAF) and the Italian National Institute for Nuclear Physics (INFN). The investigation is supported by the ASI grant  I/028/12/6. We thank the ASI management for unfailing support during AGILE operations. We acknowledge the effort of ASI and industry personnel in operating the  ASI ground station in Malindi (Kenya), and the data processing done at the ASI/SSDC in Rome: the success of AGILE scientific operations depends on the effectiveness of the data flow from Kenya to SSDC and the data analysis and software management.

We gratefully acknowledge financial support from the agencies and organizations of the CTA Observatory \url{http://www.cta-observatory.org/consortium}.

This accepted version of the manuscript is released under this license \ccbyncnd. The published version is available here \url{https://doi.org/10.1016/j.ascom.2022.100570}.




\bibliographystyle{cas-model2-names}

\bibliography{cas-refs}

\end{document}